# LARGE SCALE DISTRIBUTION OF MATTER IN THE NEARBY UNIVERSE AND ULTRA-HIGH ENERGY COSMIC RAYS

Gustavo. A. Medina Tanco[1,2]

[1]*Royal Greenwich Observatory, Madingley Rd., Cambridge CB3 0EZ, UK*

[2]*Instituto Astronômico e Geofísico, Universidade de São Paulo, Brasil*

e-mail: gmt@ast.cam.ac.uk

## ABSTRACT

The deflection of extragalactic ultra-high energy protons ($E > 4 \times 10^{19}$ eV) travelling to our galaxy is studied assuming that visible matter traces both, the sources of the particles and the intergalactic magnetic field. The reversal scale and the magnitude of the magnetic field are determined by the local density of matter. The CfA Redshift Catalog is used to determine the distribution of galaxies inside 50 Mpc, where the sources are believed to reside. It is demonstrated that the arrival directions of ultra high energy cosmic rays are consistent with the distribution of galaxies inside 50 Mpc and that the proposed clusters of events point to regions of high density of sources instead of individual ones.

Ultra high energy cosmic rays (UHECR), and specially those above the Greisen-Zatsepin-Kuzmin (GZK) cutoff [1,2] (EHECR), are a challenge for our understanding of particle physics and of the characteristics of the nearby universe. They pose, in some respects, the same kind of problems as gamma ray bursts (GRB) do; in fact, a connection between both phenomena has been proposed (e.g., [3,4], but see [5,6]). They are relatively rare events, with no certain optical counterpart, poor angular determinations, unknown powering mechanism and arguable distance scale. Although other particles cannot be disregarded, observational evidence seems to supports a hadronic nature, being protons the currently accepted working hypothesis [7]. In this context, two conclusions can be drawn: (1) due to interactions with the cosmic microwave background radiation (and as long as the relativity principle holds for Lorentz factors of $\approx 10^{11}$ [8]) the sources cannot be farther than $\approx 50$ Mpc [9]; (2) despite the fact that their trajectories are bent by the intervening galactic and intergalactic magnetic fields, the high energies involved should make possible some kind of astronomy [6,10-13]. So far, no correlation between the arrival directions of UHECR and the galactic plane is supported by the existing data [14-15]. Therefore, the sources of the particles are spread over a larger volume being, very likely, extragalactic. At present, however, statistics are very poor, and no undisputed source identification exist. The discussion is centered on broad classes of potential sources, or large scale groupings of objects rather than on individual candidates [16-20]. The main problem lies in our lack of knowledge of the intergalactic magnetic field (IGMF), except but for some few observational determinations and upper limits [21-23] and numerical simulations of cosmological structure formation (see [18] and references there in). These constraints, however, point to an IGMF structure that follows the distribution of matter as traced by the distribution of galaxies. Therefore, a high degree of inhomogeneity can be expected, with relatively high values of $B_{IG}$ over small regions ($\approx 1$ Mpc) of high matter density (e.g., $B_{IG} \approx 3\times10^{-7}$ G for the Virgo cluster [21] or $B_{IG} \approx 10^{-6}$ G for the Coma cluster [23]), pervading vast low density/low $B_{IG}$ regions with $B_{IG} < 10^{-9}$ G. This is a fundamental property of the problem that, combined with the nearness of the sources, makes of any line of sight a special case. The calculations I present here are an attempt to characterize the angular two-dimensional distribution of the arriving particles, taking into account the inhomogeneity of $B_{IG}$ and of the distribution of the potential sources of UHECR in a fully three-dimensional scheme. The results are displayed in the form of all-sky images of the celestial sphere, as should be seen by an UHECR detector.

It is assumed that the UHECR are protons, and that their sources are extragalactic and hosted by, or associated with, normal galaxies. It is further assumed that the magnetic field scales as $n_{gal}^{2/3}$, where $n_{gal}(\mathbf{r})$ is the local density of galaxies as derived from the CfA Redshift Catalog [24]. The IGMF is considered to be organized in cells of homogeneous field, such that the orientation of $\mathbf{B}_{IGM}$ of adjacent cells is uncorrelated. The size of each cell is approximated by the reversal scale, $L_c$, which relates to the IGMF through the expression $L_c \propto [B_{IGM}(\mathbf{r})]^{-2}$, and the normalization condition $L_c = 1$ Mpc for $B_{IGM} = 10^{-9}$ G is adopted (c.f., [22]). No attempt is made to include the galactic magnetic field, but its effect is assessed in several works [6,10-12,25]

Three dimensional simulations of UHECR propagating non-diffusively in the above scenario are carried out for different conditions. Only protons with $E > 4 \times 10^{19}$ eV are considered and, therefore, the most relevant energy loss mechanism is photomeson production by p-γ interactions with the cosmic microwave background radiation field [26]. This is included as in [6].

In figure 1, the two-dimensional projection in galactic coordinates (*l,b*), of the distribution of known galaxies [24] with radial distance d ≤ 50 Mpc (radial velocity v(He) ≤ 2500 km/sec) is plotted. It is easy to appreciate the inhomogeneity of the distribution. The supergalactic plane (SGP) is visible as the elongated, bent clustering of objects running from south to north between $l \approx 0°$ and $\approx -110°$. It is important to note that there is no clear signature of the SGP between $l \approx 0°$ and $l \approx 180°$ when only galaxies inside the canonical r ≤ 50 Mpc volume are considered (that is, where the UHECR sources are thought to reside). Consequently, if any strong correlation is found between UHECR and the SGP in that region of the sky, they must come from r > 50 Mpc, and either the particles are probably neutral or the significance of GZK cutoff should be reevaluated. At present it is not clear whether such correlation actually exist [14,16,18-19,27]. The lack of objects in the central strip in the figure at $b \approx \pm 10°$ is only the result of the obscuration produced by the galactic plane, and does not represent a real depletion in the galaxy distribution.

In figure 2 we show numerical results for UHECR injected with a monochromatic energy spectrum at the location of the galaxies (see figure 1). The particle fluxes are diluted according to a $r^{-2}$ law and the luminosity in UHECR, $L_{CR}$, is considered constant for every galaxy. This standard candle approximation corresponds, for example, to a scenario in which the acceleration sites can develop in (or be associated with) any kind of galaxy with the same probability, but their lifetimes are smaller than their recurrence period so that only one acceleration site exist at a time in a given galaxy. For example, a short lived, non-recurrent stage in the life of a compact object could reproduce this behavior. Calculations have also been carried out for the case in which the luminosity in UHECR scales with the absolute luminosity of the galaxy, i.e., $L_{CR} \propto L_{gal}(B)$. The results are qualitatively similar (cf. figure 3) and will be presented elsewhere in detail [28].

The contours in figure 2 represent the two-dimensional arrival distribution function of UHECR protons for any direction of the sky calculated for a monoenergetic injection at $E_{inj}$ = $3 \times 10^{20}$ eV, and IGMF normalized by its value at the Virgo cluster, $B_{IG} \approx 10^{-7}$ G [21]. The squares are the positions of the 12 Yakutsk events with energies > $4 \times 10^{19}$ eV for the whole operation period up to May 1996 [29]; the circles are the 36 events from AGASA with E > $4 \times 10^{19}$ eV (until 1995.1) [14]; the triangles are 6 events from Volcano Ranch with E > $10^{19}$ eV [30]; and the hexagons are the EHECR events with E > $10^{20}$ eV [31]. The hatched strip at the middle of the figure corresponds to the region of the sky obscured by the galactic plane and where the information on extragalactic objects is incomplete. The two sets of lines demarcate the region of the sky where 90% of the events were detected by AGASA (broken line) and Yakutsk (broken-dotted line) respectively (adapted from [32]) and are representative of the sensibility of those experiments.

The model used to construct figure 3 is similar to that of figure 2, but $L_{CR} \propto L_{gal}(B)$. Four possible clusters of UHECR events, proposed by [14], are shown.

The main conclusions can be summarized as follows:

1) There is no clear signature of the SGP in the range $l \approx 0 - 180°$ when only galaxies at r < 50 Mpc are considered (see figure 1). Therefore, no correlation should be expected between the arrival directions of UHECR and the SGP in that direction of the sky unless the GZK is, for some reason, not verified.

2) The probability distribution does not change very much when either the injection energy or the assumption about the luminosity of the sources in UHECR is changed (compare figures 2 and 3 and see [28] for further details).

3) Given the non uniform sensibility of the experiments in declination, the UHECR events seem to follow reasonably well, given the low number statistics, the calculated 2D arrival distribution function, and so their sources may well be associated with normal galaxies, or some population that follows closely the distribution of visible matter in the nearby universe (d < 50 Mpc) and is well sampled by the CfA Redshift Catalog. Much the same applies to the events with E > 100 EeV.

4) From figure 3, it can be seen that one of the three pairs of events [14] ($l \cong 144°$, $b \cong 56°$) is located exactly at a maximum of the arrival probability distribution , and therefore it may not necessarily represent a pair of events coming from the same source. Something similar can be said of the other two pairs, which may be associated with some local maximum of the probability distribution. The conclusion of clustering may be, therefore, biased by the comparison of the observed distribution of the events with an isotropic distribution of sources (see [28,33] for further details). The addition of the Haverah Park event at $l$=134.1° $b$=-41° to pair number 1, makes of this *triplet* a more promising candidate for a true single UHECR source. Nevertheless, this picture is further complicated by the non-uniform sensibility of the experiments (see figure 2 for AGASA and Yakutsk, and [32] for Haverah Park and Fly's Eye), that can hide events associated with the maximum of the arrival distribution at $l \cong 35°$ $b \cong -55°$.

5) Three of the EHECR events and another 8 UHECR events from various experiments are located so unfortunately near the galactic plane that no firm conclusion can be extract regarding them. Pair 4 in figure 3 falls in this category.

The author wish to thank H. Arp for valuable comments, E. M. de Gouveia Dal Pino and J. E. Horvath for stimulating discussions, J. Cronin and M. Nagano for kindly supplying data on UHECR and J. F. Valdés-Galicia for his hospitality at the Instituto de Geofísica of the UNAM, Mexico, where this work was partially done. This work was supported by the Brazilian agency FAPESP.

**Figure Captions**

**Figure 1**: Two-dimensional projection, in galactic coordinates, of the distribution of known galaxies [24] with radial distance d ≤ 50 Mpc (radial velocity v(He) ≤ 2500 km/sec).

**Figure 2**: Arrival distribution function of UHECR protons (contours), observed UHECR events for AGASA (circles), Yakutsk (squares) and Volcano Ranch (triangles), and EHECR with E > 100 EeV (hexagons). The hatched strip at b = ± 10° corresponds to the region obscured by the galactic plane where our results are not valid. The two sets of lines demarcate the region of the sky where 90% of the events were detected by AGASA (broken line) and Yakutsk (broken-dotted line) respectively and are representative of the sensibility of those experiments [32].

**Figure 3**: Same as figure 2, but for $L_{CR} \propto L_{gal}(B)$. Only the clusters of UHECR proposed by [14] are shown. Note that they are associated with local enhancements of the calculated arrival distribution, exception made of pair 4 that is inside the galactic plane. The small dots are the galaxies inside 50 Mpc (as in figure 1).

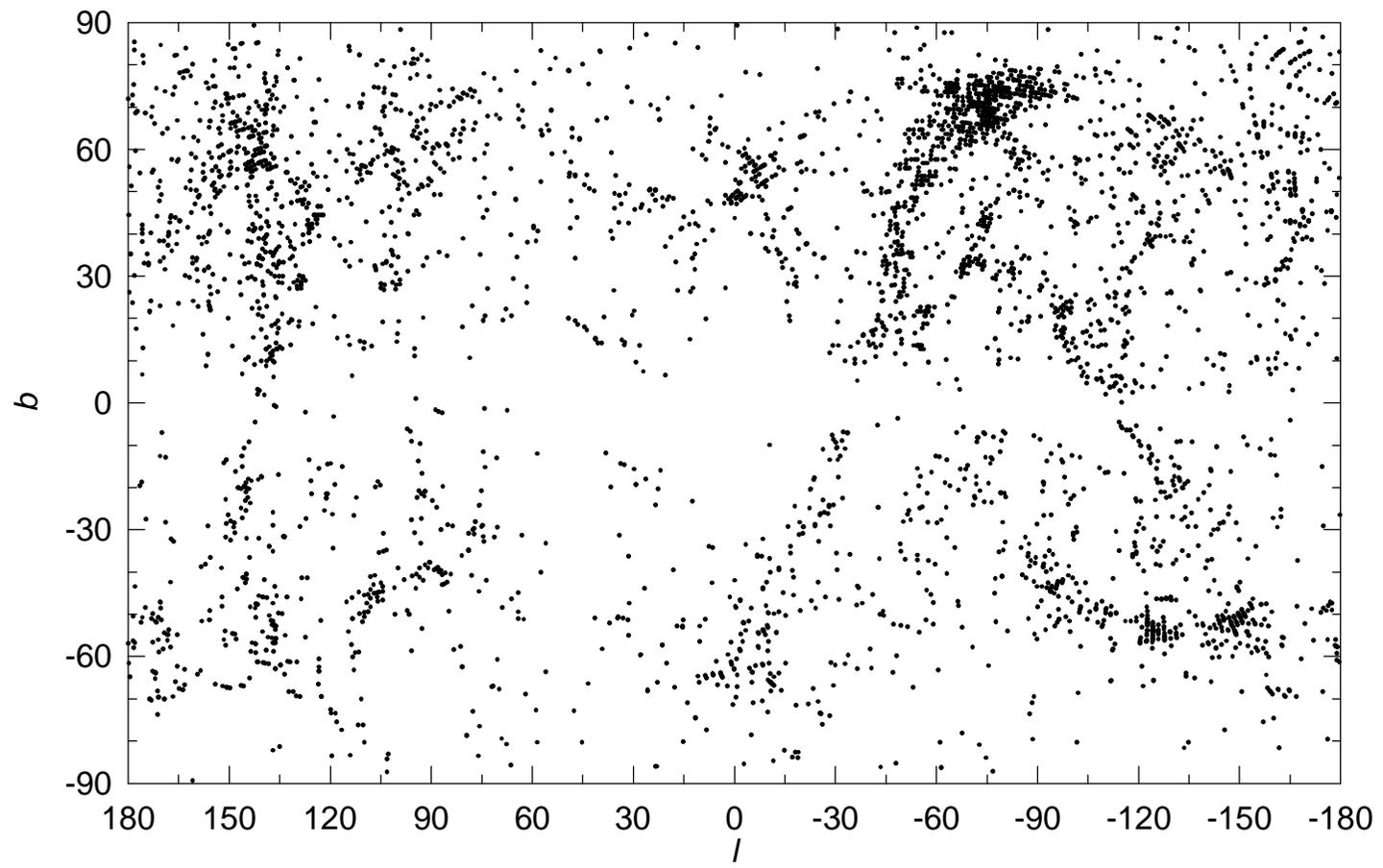

**Figure 1**

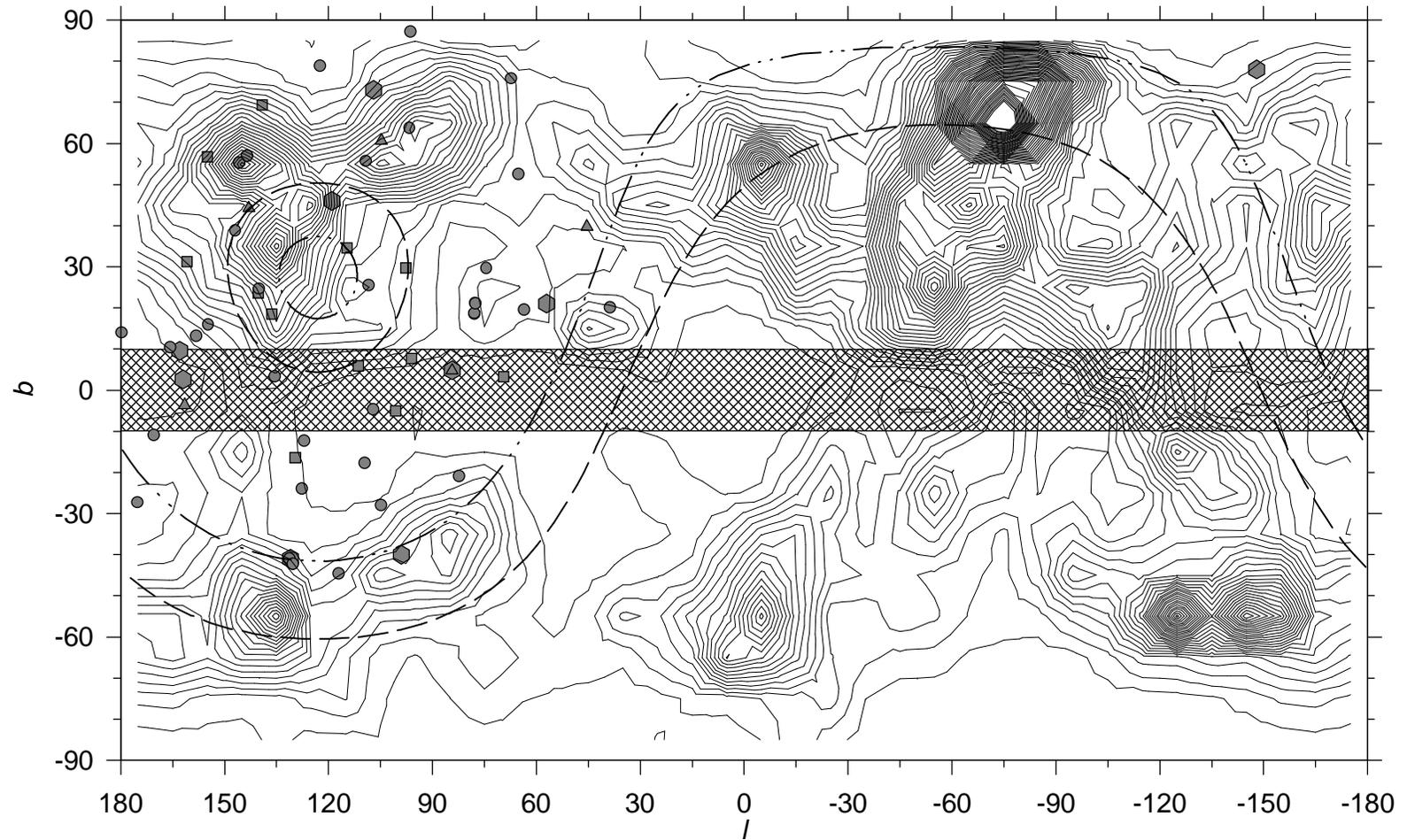

**Figure 2**

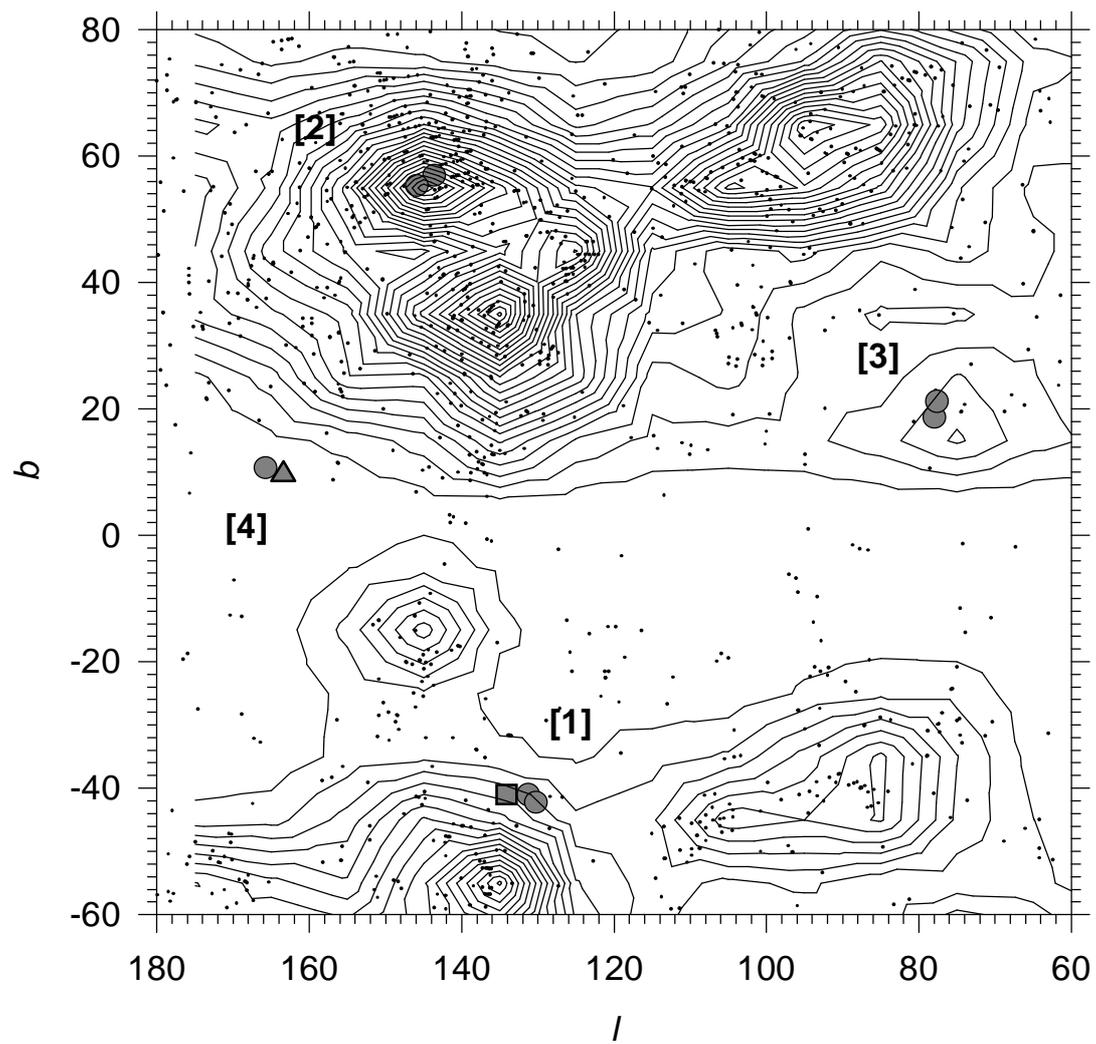

**Figure 3**